\documentclass[prb,floatfix,twocolumn,showpacs]{revtex4-1}
\usepackage{graphicx}

\usepackage[usenames,dvipsnames]{color}

\begin{document}
\title{Signatures of the many-body localization transition in the dynamics \\ of entanglement and bipartite fluctuations}
\author{Rajeev Singh, Jens H. Bardarson, Frank Pollmann}
\affiliation{Max-Planck-Institut f\"{u}r Physik komplexer Systeme, 01187 Dresden, Germany.}
\date{\today}
\begin{abstract}
The many-body localization transition is a dynamical quantum phase transition between a localized and an extended phase.
We study this transition in the XXZ model with disordered magnetic field and focus on the time evolution following a global quantum quench.  
While the dynamics of the  bipartite entanglement and spin fluctuations are already known to provide insights into the nature of the many-body localized phases, we discuss the relevance of these quantities in the context of the localization transition.
In particular, we observe that near the transition the long time limits of both quantities show behavior similar to divergent thermodynamic fluctuations.
\end{abstract}
\maketitle
\section{Introduction}
Many-body localization (MBL) occurs when Anderson localization \cite{Anderson1958} persists in the presence of interactions. 
In the pioneering work of Basko, Aleiner and Altshuler,\cite{Basko2006} the localized phase was shown to be perturbatively stable to small interactions.
This work quickly opened up a new field and many intriguing properties of this new phase were explored:
(i) due to the lack of transport, MBL systems do not thermalize,\cite{Nandkishore2014}
(ii) at finite energy densities the localization of domain walls allows stabilizing quantum and topological order which would otherwise melt,\cite{Huse2013,Kjall2014}
and (iii) following a global quantum quench, MBL phases have a characteristic logarithmic growth of entanglement as a function of time.\cite{Znidaric2008,Bardarson2012,Serbyn2013,Vosk2013,Nanduri2014}
On the experimental side, first progress has been made in realizing such systems:
In Ref.~\onlinecite{Schreiber2015} the effect of localization was observed in a cold atom experiment where a charge density wave failed to relax in the localized phase.
By measuring $I-V$ characteristics of amorphous iridium-oxide, Ref.~\onlinecite{Ovadia2014} has provided evidence for a finite temperature insulator where the MBL mechanism might be at play.

As the MBL transition occurs in eigenstates at finite energy densities instead of just the ground state, this transition is a dynamical quantum phase transition.~\cite{Pal2010}
Many aspects of this transition from an MBL phase to an extended one are still not fully understood. 
An interesting feature of the transition is that, in principle, the critical disorder strength depends on energy density, yielding a so-called \emph{many-body mobility edge}.~\cite{Basko2006,Huse2013,Kjall2014,Luitz2014,Bera2015}
Novel real space renormalization group methods have been developed in which this transition is given by an infinite randomness RG fixed point.~\cite{Vosk2013,Pekker2013a,Vosk2014a,Potter2015}
In this work, we consider the anti-ferromagnetic spin-1/2 XXZ chain and study the time evolution of the entanglement as well as the bipartite fluctuations following a global quench.
We focus on the evolution of the probability distribution and show that the standard deviation of the long-time limit can be used to detect the MBL transition. 
The observations made for the bipartite fluctuations are particularly useful for an experimental detection of the transition in cold atomic systems.
We furthermore discuss the behavior after a very long time following the global quench in comparison to that in the thermal state and in the diagonal ensemble.
This paper is organized as follows: In section \ref{model} we describe the model and briefly mention some of its properties.
In section \ref{QuenchDynamics} we describe the global quench protocol followed by the description of the behavior of entanglement and bipartite fluctuations.
We then compare the long time behavior to that in the thermal state and diagonal ensemble.
We present the quench results for the non-interacting case next and finally conclude by providing a summary and outlook in section \ref{Conclusions}.

\section{Model} \label{model}
\begin{figure}
\begin{center}
\includegraphics[clip,trim=40mm 165mm 40mm 0mm,width=\linewidth]{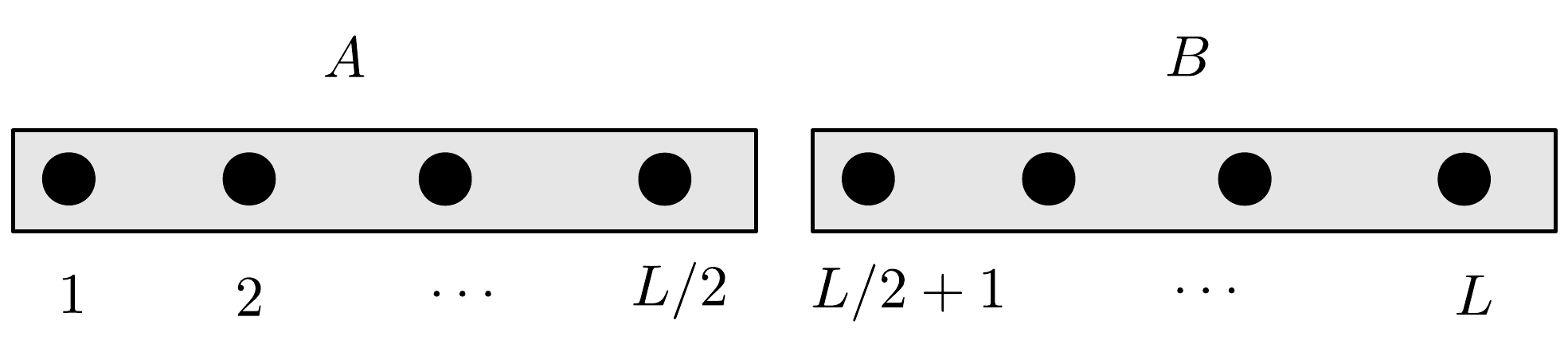}
\end{center}
\caption{A schematic representation of a 1D spin chain subdivided into two parts $A$ and $B$, which are used to calculate the entanglement entropy and bipartite fluctuations.}
\label{Cartoon}
\end{figure}
We consider the anti-ferromagnetic spin-1/2 XXZ model on a one-dimensional chain in the presence of a disordered $z$-directed magnetic field. 
The Hamiltonian is given by
\begin{equation} \label{Hamiltonian}
H = J \sum_{i=1}^{L-1} (S^x_i S^x_{i+1} + S^y_i S^y_{i+1}+
  \Delta S^z_i S^z_{i+1}) + \sum_{i=1}^L h_i S^z_i,
\end{equation}
where $J>0$ is the anti-ferromagnetic coupling strength between neighboring spins, $\Delta$ is the anisotropy parameter and $h_i$'s are the uncorrelated random external fields. 
Throughout this paper we consider the case $J=\Delta=1$ (except for the non-interacting case when $\Delta=0$) and choose $h_i$  from a uniform distribution $[-\eta,\eta]$.
\footnote{Note that this model can be mapped to a model of interacting spinless fermions in one-dimension via the Jordan-Wigner transformation.}
The Hamiltonian~(\ref{Hamiltonian}) provides a simple model to study the MBL phenomena numerically.~\cite{Chiara2006,Znidaric2008,Pal2010,Bardarson2012,Luitz2014}
This model shows a localization transition at $\eta_c \approx 3.6$ at infinite temperature corresponding to eigenstates in the middle of the spectrum.~\cite{Pal2010,Luitz2014}
It has been argued that MBL systems have a many-body mobility edge,~\cite{Huse2013} which was first observed numerically in transverse field Ising chain.~\cite{Kjall2014}
The mobility edge has also been obtained for the XXZ chain in Ref.~\onlinecite{Luitz2014} and spinless fermions in Ref.~\onlinecite{Bera2015}.

\section{Quench Dynamics} \label{QuenchDynamics}
Following Refs.~\onlinecite{Znidaric2008,Bardarson2012} we consider a global quench starting from a simple product state.
In particular, we choose the N\'eel state (a product state of alternating up and down spins) as the initial state and study the time evolution of the system using exact diagonalization.
This simulation corresponds to a global, sudden quench in which we start from the ground state of Hamiltonian (\ref{Hamiltonian}) with an infinite staggered field which is then turned off at $t=0$.
In the following, we perform a detailed analysis of the time evolution of entanglement and bipartite fluctuations.
\subsection{Entanglement entropy} \label{EntanglementSection}
\begin{figure}
\begin{center}
\includegraphics[width=0.99\linewidth]{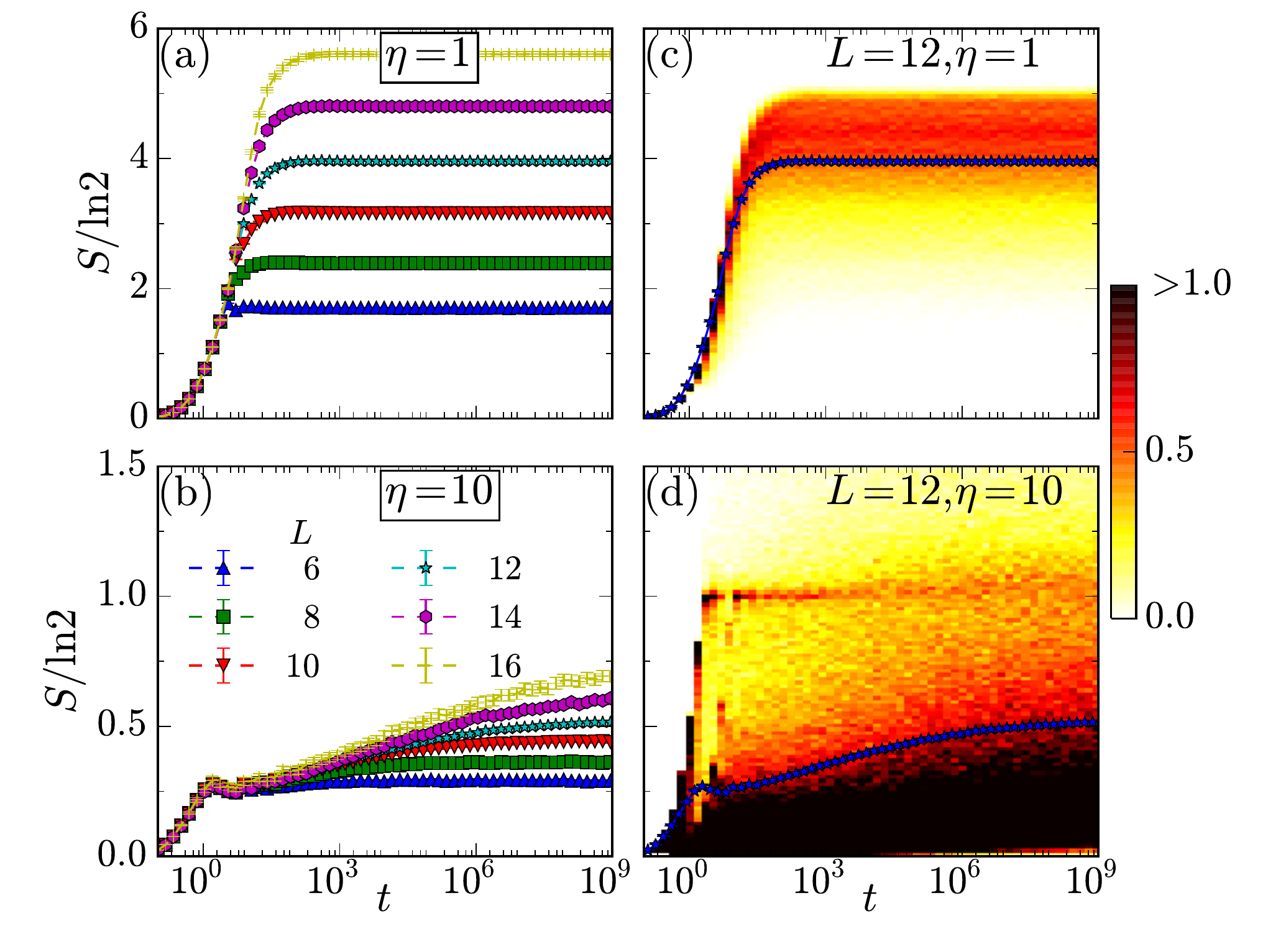}
\end{center}
\caption{(color online).
(a-b) Time evolution of the entanglement entropy averaged over disorder realizations, for different disorder strengths $\eta = 1, 10$ and system-sizes $L$.
The mean entanglement saturates after an initial growth for both weak (a) and strong disorder (b).
In the case of strong disorder it shows a logarithmic growth over several decades before saturating to a much lower value compared to the weak disorder case.
To highlight the qualitative difference between strong and weak disorder we also show the evolution of the distribution of entanglement, color scale, for $L = 12$ in (c, d).
}

\label{ent}
\end{figure}
\begin{figure}
\begin{center}
\includegraphics[clip,trim=0mm 0mm 0mm 0mm,width=0.99\linewidth]{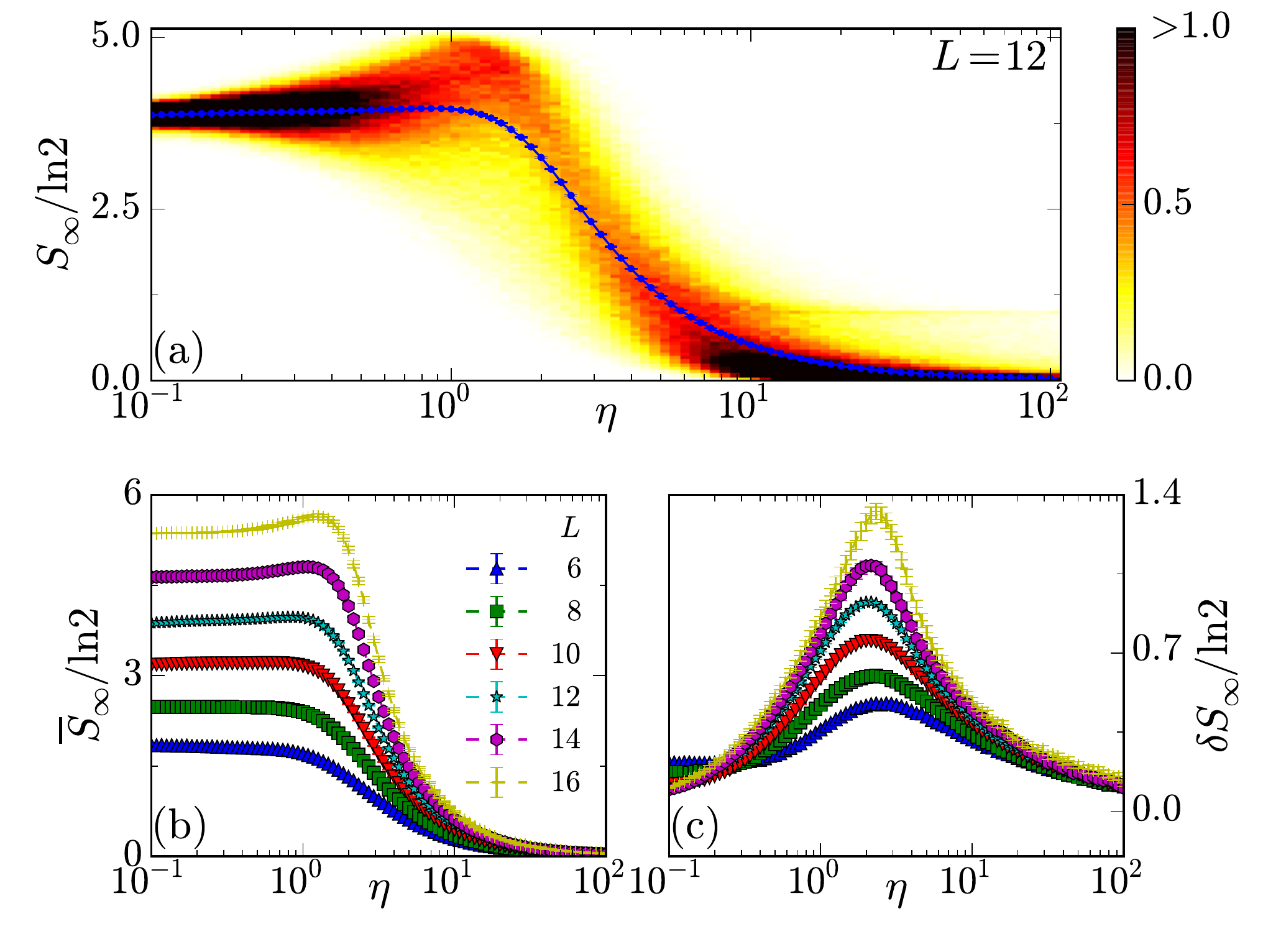}
\end{center}
\caption{(color online).
  (a) The change in the distribution of saturation entanglement $S_{\infty}$ ($t_{\infty}=10^{16}$) as a function of disorder strength $\eta$.
The mean (b) and standard deviation (c) of the saturation entanglement entropy as a function of disorder strength for different system sizes.
The standard deviation behaves like divergent thermodynamic fluctuations showing a peak which becomes higher with system size.
}
\label{ent_sat}
\end{figure}
We start by considering the evolution of entanglement between two equal partitions of the chain (Fig.~\ref{Cartoon}).
For pure states, the entanglement entropy is given by the von Neumann entropy of the reduced density matrix $\rho$ corresponding to either subsystem.
The reduced density matrix of the left half of the chain ($A$) for the state $|\psi \rangle$, is $\rho_{_A} = \text{Tr}_{_B} (|\psi \rangle \langle \psi|)$, where we have traced over the degrees of freedom of the right-half of the chain ($B$).
The von Neumann entropy of the state $\rho_{_A}$ is then given by
$$S = -\text{Tr}_{_A} \rho_{_A} \ln \rho_{_A},$$
where the trace is now over the remaining degrees of freedom ($A$).
Figure~\ref{ent}~(a-b) show the time evolution of $S$ averaged over disorder realizations for different system sizes ($L=6,\cdots,16$) and two different disorder strengths ($\eta=1,10$).
We have used $10^4$ disorder realizations for $L\le12$, $10^3$ for $L=14$ and $500$ for $L=16$.
For weak disorder ($\eta=1$) the entanglement shows a fast linear growth which then rapidly saturates to a value $S_{\infty}$.
This linear growth is due to the spreading of correlations at a finite speed before saturating because of finite size of the system.~\cite{Lieb1972,Calabrese2006}
The saturation value follows a volume law $S_{\infty} = \alpha L$ with $\alpha$ being close to its maximum value of $\alpha_{\mathrm{max}}=\ln(2)/2$ (for a partition of the system into two half chains).
For strong disorder ($\eta=10$), the  system shows a rapid linear growth only for a short duration.\cite{Znidaric2008,Bardarson2012}
Then the localization causes the linear growth to terminate and is followed by a slow logarithmic growth for a long time before it eventually saturates for finite systems.
The saturation value $S_{\infty}$  still shows a volume law but the coefficient $\alpha$ is much smaller than it is in the weak disorder case.
We note that the duration of logarithmic growth increases with increasing disorder and system size, and we have chosen large enough time interval to study the saturation properties for our system sizes.
This logarithmic growth in the localized phase has been studied in detail in recent works~\cite{Znidaric2008,Bardarson2012,Serbyn2013,Vosk2013,Nanduri2014} and has been explained via an interaction induced dephasing mechanism.~\cite{Serbyn2013,Vosk2013}
To gain further insight into the details of entanglement dynamics, we plot the distribution of $S$ in Fig.~\ref{ent}~(c-d) at different disorder strengths for $L=12$, and find that it differs strongly between the two cases.
In the case of weak disorder there is a single peak which broadens and shifts to higher value of $S$ before saturating.
For the strongly disordered case, in contrast, the starting distribution with a single peak splits into a bimodal distribution with two peaks at intermediate times; the larger peak being near zero while the smaller and much sharper peak is at $\ln(2)$.
This value of the second peak corresponds to cutting one singlet in the partition between left and right half of the chain.~\cite{Laflorencie2005,Bauer2013}
The first peak at smaller value of $S$ slowly broadens and shifts to slightly larger values before its tail merges with the second smaller peak and the asymptotic distribution thus has a single peak. 
During this broadening the second peaks stays at $\ln(2)$.
The time interval over which this broadening of the main peak happens corresponds exactly to the duration of logarithmic growth.
Figure~\ref{ent_sat}(a) shows the asymptotic distribution of entanglement as a function of disorder strength.
For very weak disorder the entropy distribution is centered around a relatively high value.
The distribution first broadens with disorder followed by the main peak shifting to lower values.
Both for very weak and for strong disorder, the distribution is very narrow. 
To make these observations more quantitative, we also show the mean $\overline{S}_{\infty}$ and standard deviation $\delta S_{\infty}$ of $S_{\infty}$ in Fig.~\ref{ent_sat}~(b-c).
The mean changes slowly for weak disorder but suddenly decreases to very small values as the disorder is increased beyond a critical value.
The standard deviation behaves similar to fluctuations in thermodynamic phase transitions and shows a peak which gets higher with system size.
The divergence of the standard deviation is due to the fact that near the transition small changes in the energy densities and disorder realizations decide whether the system is localized or extended. 
Thus this quantity is a good observable to pinpoint the transition.

\subsection{Bipartite fluctuations} \label{BipartiteSection}
\begin{figure}
\begin{center}
\includegraphics[width=0.99\linewidth]{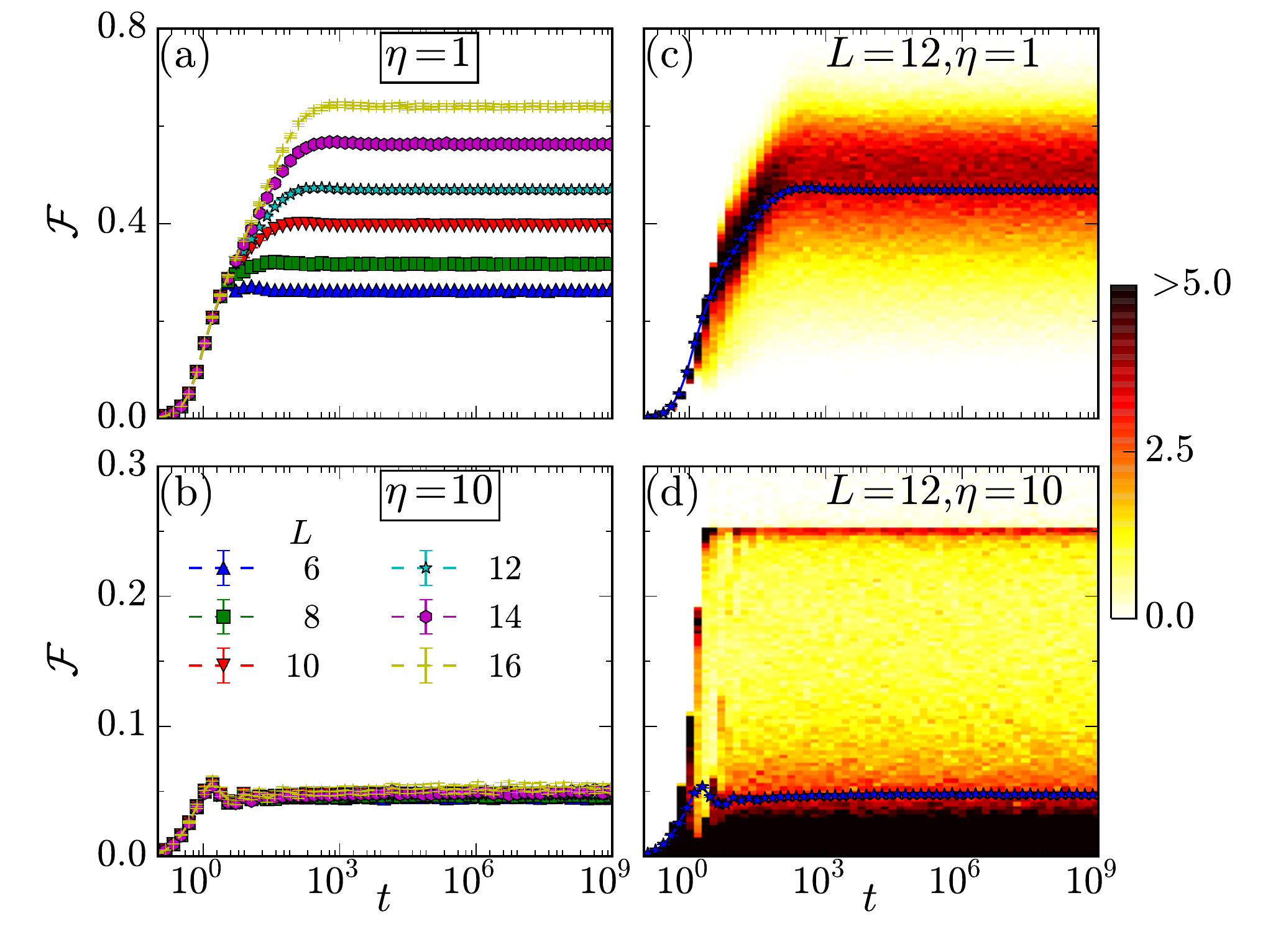}
\end{center}
\caption{(color online).
(a-b) Time evolution of bipartite fluctuations averaged over disorder realizations, for different disorder strengths $\eta = 1, 10$ and system-sizes $L$.
Just like the entanglement, the mean of bipartite fluctuations saturates after an initial growth for both weak (a) and strong disorder (b).
However unlike the entanglement entropy it does not show a logarithmic growth for strong disorder.
We show the evolution of the distribution of bipartite fluctuations, color scale, for $L = 12$ in (c, d).
}
\label{bpf}
\end{figure}
\begin{figure}
\begin{center}
\includegraphics[width=0.99\linewidth]{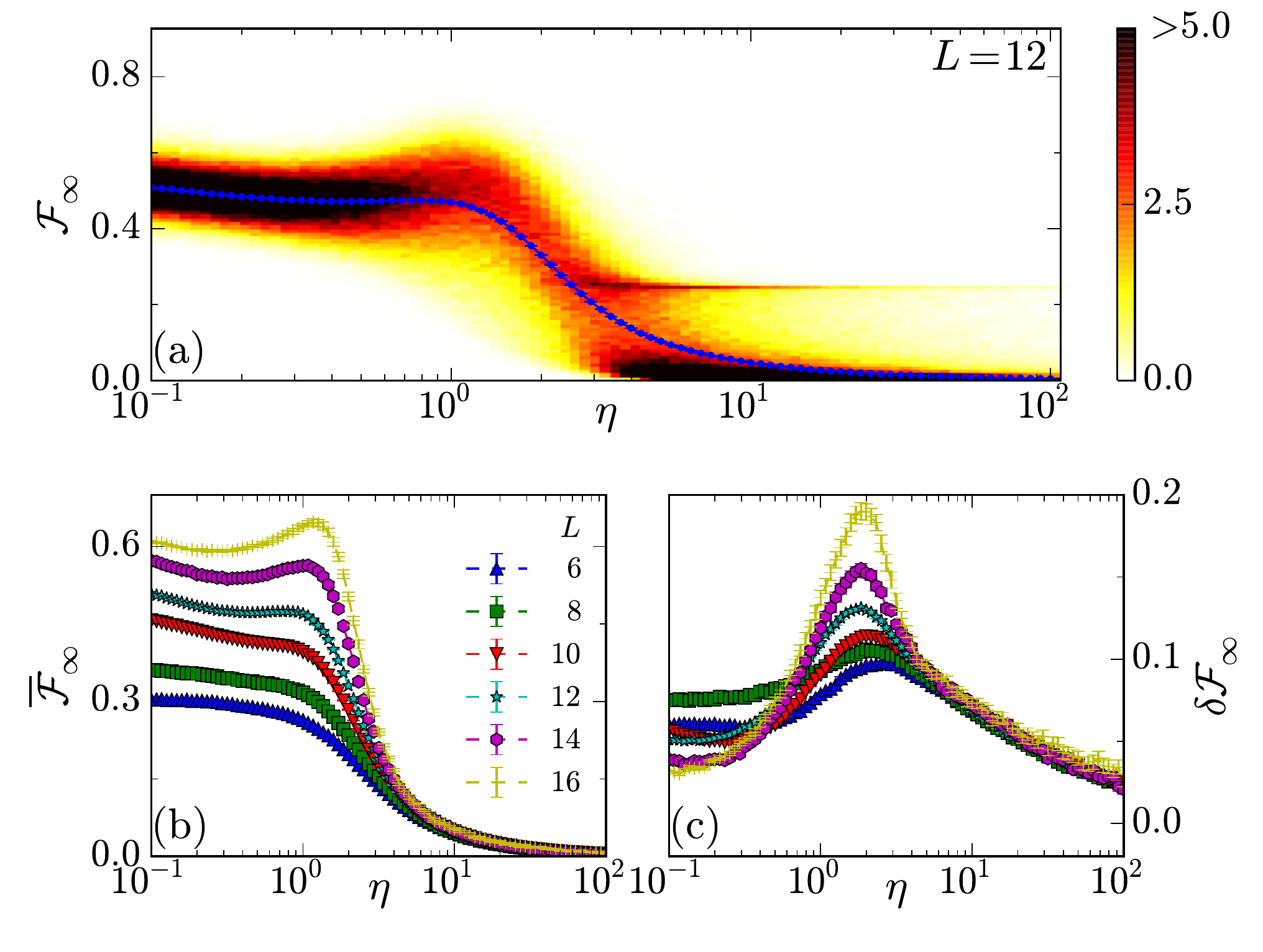}
\end{center}
\caption{(color online).
(a) The distribution of bipartite fluctuations at long times ($t_{\infty}=10^{16}$) as a function of disorder strength clearly shows the appearance of second sharp peak at $1/4$ near the expected transition.
The weight of the second peak gradually shifts to the main peak with increasing disorder, resulting in a decrease of the mean.
We show the mean (b) and standard deviation (c) of bipartite fluctuations with disorder strength for different system sizes.
}
\label{bpf_sat}
\end{figure}
We now consider the dynamics of bipartite fluctuations~\cite{Rachel2012} following the global quench.
While the total magnetization $\hat{S}^z_{\mathrm{total}}$ of Hamiltonian (\ref{Hamiltonian}) is conserved, the magnetization of the half-chain
$$\hat{S}^z_{L/2} = \sum_{i=1}^{L/2} \hat{S}^z_i,$$ 
fluctuates. 
We define the bipartite fluctuations $\mathcal{F}$ as the quantum fluctuations of $\hat{S}^z_{L/2}$,
$$\mathcal{F} \equiv \langle \psi | (\hat{S}^z_{L/2})^2 | \psi \rangle
     - \langle \psi |  \hat{S}^z_{L/2}    | \psi \rangle^2.$$
Entanglement is difficult to measure experimentally though there have been recent suggestions to observe its effects in MBL systems.~\cite{Vasseur2014,Ho2015}
On the other hand $\mathcal{F}$ can be accessed in experiments as follows.
There is an exact mapping between Hamiltonian~(\ref{Hamiltonian}) and hardcore bosons on a 1D lattice, and the bipartite fluctuations defined above are equivalent to particle number fluctuations in the bosonic system, which can be measured using single atom microscopy.~\cite{Bakr2009,Sherson2010}
Figure~\ref{bpf}~(a-b) shows the time evolution of the disorder averaged $\mathcal{F}$.
The mean $\mathcal{F}$ grows rapidly and saturates at very short time scale both in the extended and localized phases.
Unlike entanglement there is no logarithmic growth for strong disorder and the bipartite fluctuations saturate to a much smaller value almost independent of system size.\cite{Bardarson2012}
We present the distribution of $\mathcal{F}$ as a function of time for weak and strong disorder in Fig.~\ref{bpf}~(c-d).
The behavior at weak disorder strength for both the short and long time limit is very similar to that of entanglement. 
In particular, we obtain a  peak that broadens as a function of time until saturation.
However we can clearly see the difference for strong disorder.
The short time evolution is qualitatively similar to entanglement, in that $\mathcal{F}$ also shows a bimodal distribution with a second peak at $1/4$.~\footnote{{ The value $1/4$ again corresponds to cutting a singlet across the partition just like entanglement and is easy to understand if we consider a two-site example in a singlet state: $\hat{S}^z_{L/2}$ is simply $\hat{S}^z_1$ for the first site and $(\hat{S}^z_1)^2 = 1/4$ while $\langle \psi | \hat{S}^z_1 | \psi \rangle = 0$ for the singlet state, hence we obtain $\mathcal{F} = 1/4$ if the partition cuts one singlet.}}
The long time behavior on the other hand is very different from that of entanglement.
For $\mathcal{F}$ the first bigger peak does not broaden with time and the second peak persists even after a long time.
This also corresponds to the absence of logarithmic growth and the distribution saturates much more rapidly than that of entanglement.
The absence of logarithmic growth implies that though the many-body wavefunction continues to evolve, bipartite fluctuations are not affected by the dephasing mechanism and attain their asymptotic values on a much smaller time scale.
It also implies that though $\mathcal{F}$ can distinguish between localized and extended phases, it is insensitive to the effects of interactions and hence can not distinguish MBL from Anderson localization.
The change in the saturation properties of the bipartite fluctuations ($\mathcal{F}_{\infty}$) with disorder strength also captures the transition quite well qualitatively~(Fig.~\ref{bpf_sat}).
A second peak at the value $1/4$ appears in the asymptotic distribution near the transition, see Fig.~\ref{bpf_sat}~(a).
We show the mean $\overline{\mathcal{F}}_{\infty}$ and standard deviation $\delta \mathcal{F}_{\infty}$ of $\mathcal{F}_{\infty}$ as a function of disorder strength $\eta$ in Fig.~\ref{bpf_sat}~(b-c).
In the MBL phase, the individual particles can move around only a short distance within some localization length.
Therefore the particle number fluctuations get contribution only from particles which are near the partition and as a result should be independent of system size.
We find that this is indeed the case for strong disorder strengths.
This can also be seen in Fig.~\ref{bpf}~(b) where the time-evolution of $\mathcal{F}$ is almost independent of system size.
Just as in the case of entanglement the standard deviation behaves like thermodynamic fluctuations with the peak becoming more pronounced with system size.

\subsection{Comparison to thermal state} \label{ThermalState}
\begin{figure}
\begin{center}
\includegraphics[width=0.99\linewidth]{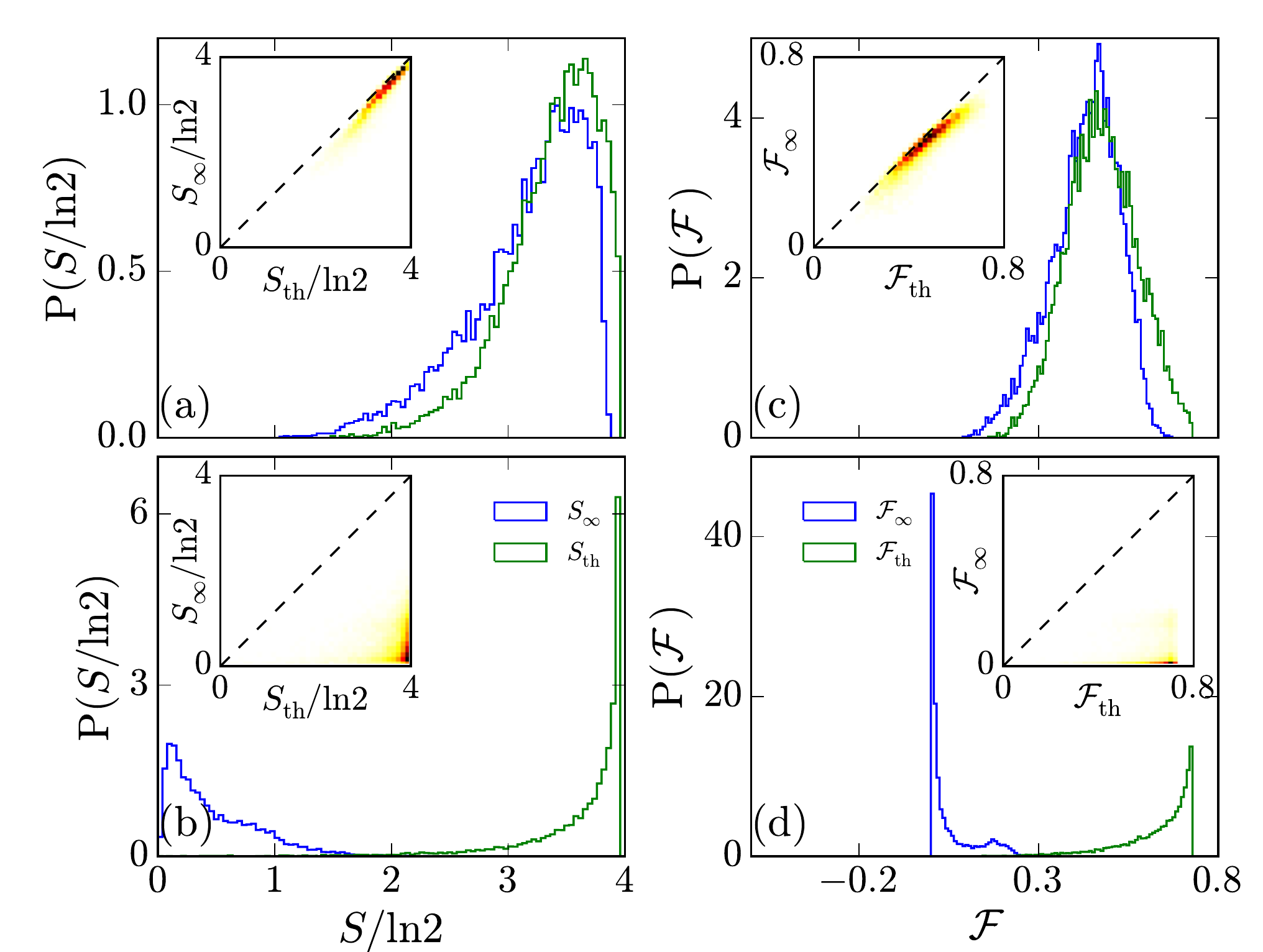}
\end{center}
\caption{(color online).
Asymptotic ($t_{\infty}=10^{16}$) entanglement and bipartite fluctuations compared with their values in the thermal state for weak ($\eta=1$ in a,c) and strong disorder ($\eta=10$ in b,d).
We use an unequal partition ($L_A=4, L_B=8$) for this comparison as the thermal state state predictions are expected to be better if the bath (here subsystem $B$) is larger than the system of interest ($A$).
Long time after the quench, both quantities agree with the value predicted from the thermal state for $L=12$ in the case of weak disorder (a,c) while the agreement is very poor at strong disorder (b,d).
We plot the 2D histograms in the insets which show the correlation between the asymptotic and thermal values in each case.
}
\label{ThermalPrediction}
\end{figure}

The notion of thermalization in a closed quantum system implies that a generic system would eventually relax and its asymptotic behavior can be described by a thermal state with the same energy density.
In the case of integrable systems, one needs to take into account all conserved quantities instead of just the energy,\cite{Rigol2007} however Hamiltonian~(\ref{Hamiltonian}) is not integrable for any finite $\eta$.
As a result, in the extended phase we expect that the asymptotic properties can be understood from thermalization, i.e., all properties depend only on the energy density of the initial state.
We compare the properties of our system at long times with the thermal state $\rho_{_{\mathrm{th}}} \propto e^{-\beta H}$ with $\beta$ chosen such that $\langle \psi_0 | H | \psi_0 \rangle = \mathrm{Tr}(\rho_{_{\mathrm{th}}} H)$.
The thermalization arguments are applicable if the subsystem of interest (say $A$) is much smaller than the rest of the system ($B$) as in such situations the subsystem $B$ acts as a bath for the subsystem $A$.
Thus we use an unequal partition ($L_A=4, L_B=8$) to compare the asymptotic entanglement and the thermal entropy.
For weak disorder ($\eta=1$) the system is in an extended phase and we find a good agreement between the thermal and asymptotic values for both entropy and bipartite fluctuations as seen in Fig.~\ref{ThermalPrediction}~(a,c) and their insets.
For strong disorder we show the asymptotic and thermal values in Fig.~\ref{ThermalPrediction}~(b,d) and find a very poor correlation.
Whereas the thermal values for a given energy density is still predicted to be large, both quantities saturate to much smaller value in each realization.
This is a very clear signature of failure of thermalization for localized systems.

\subsection{Comparison to the diagonal ensemble} \label{DiagonalEnsemble}
\begin{figure}
\begin{center}
\includegraphics[width=0.99\linewidth]{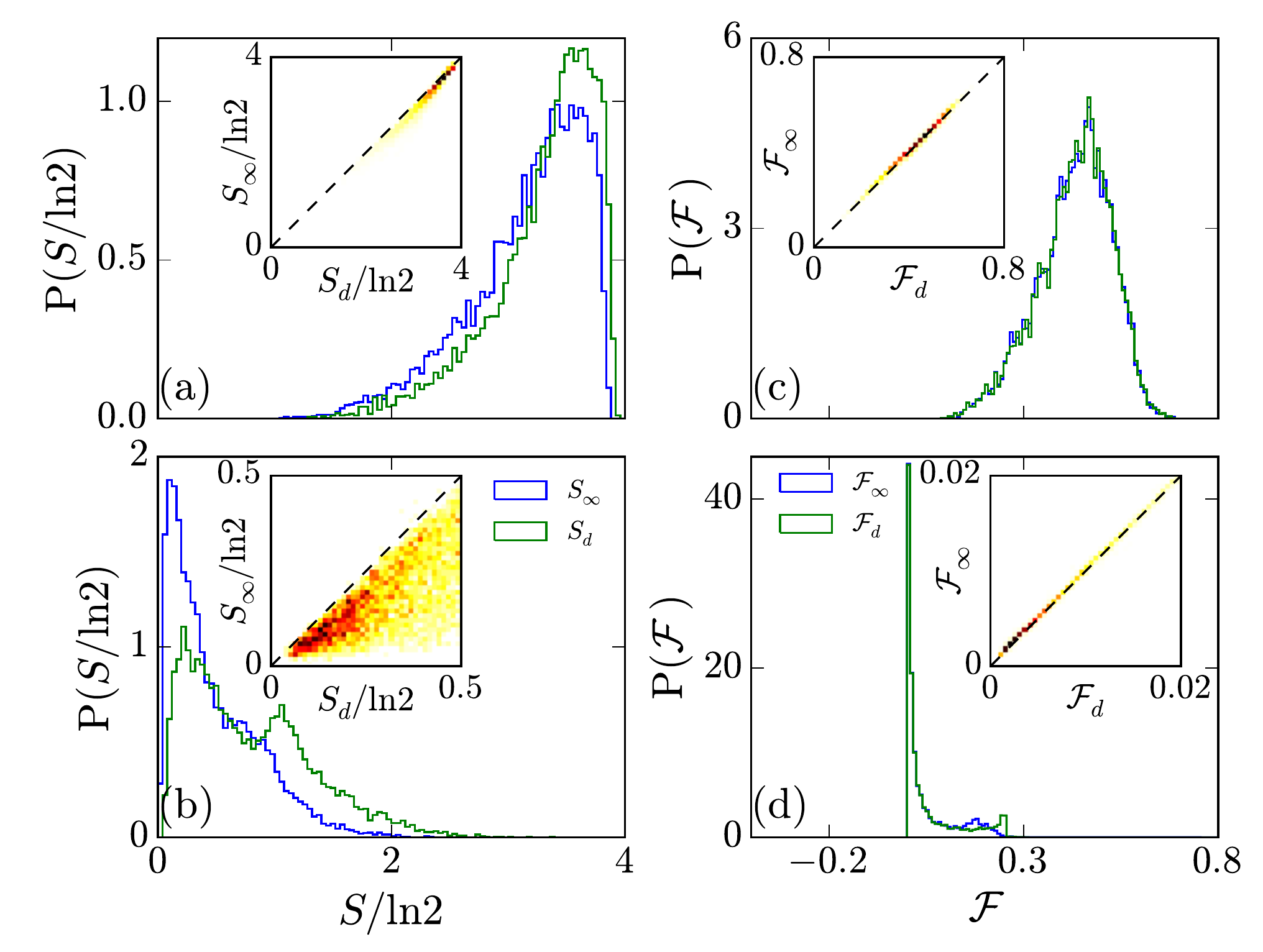}
\end{center}
\caption{(color online).
Asymptotic ($t=10^{16}$) entanglement and bipartite fluctuations compared with their values in the diagonal ensemble for weak ($\eta=1$ in a,c) and strong disorder ($\eta=10$ in b,d).
We use an unequal partition ($L_A=4, L_B=8$).
Time averaged entanglement long time after the quench agrees with the value predicted from the diagonal ensemble for $L=12$ in the case of weak disorder (a) while the agreement is not good at strong disorder (b).
(c-d) The prediction of bipartite fluctuations on the other hand is very good for both weak and strong disorder.
}
\label{DiagonalPrediction}
\end{figure}
\begin{figure}
\begin{center}
\includegraphics[width=0.99\linewidth]{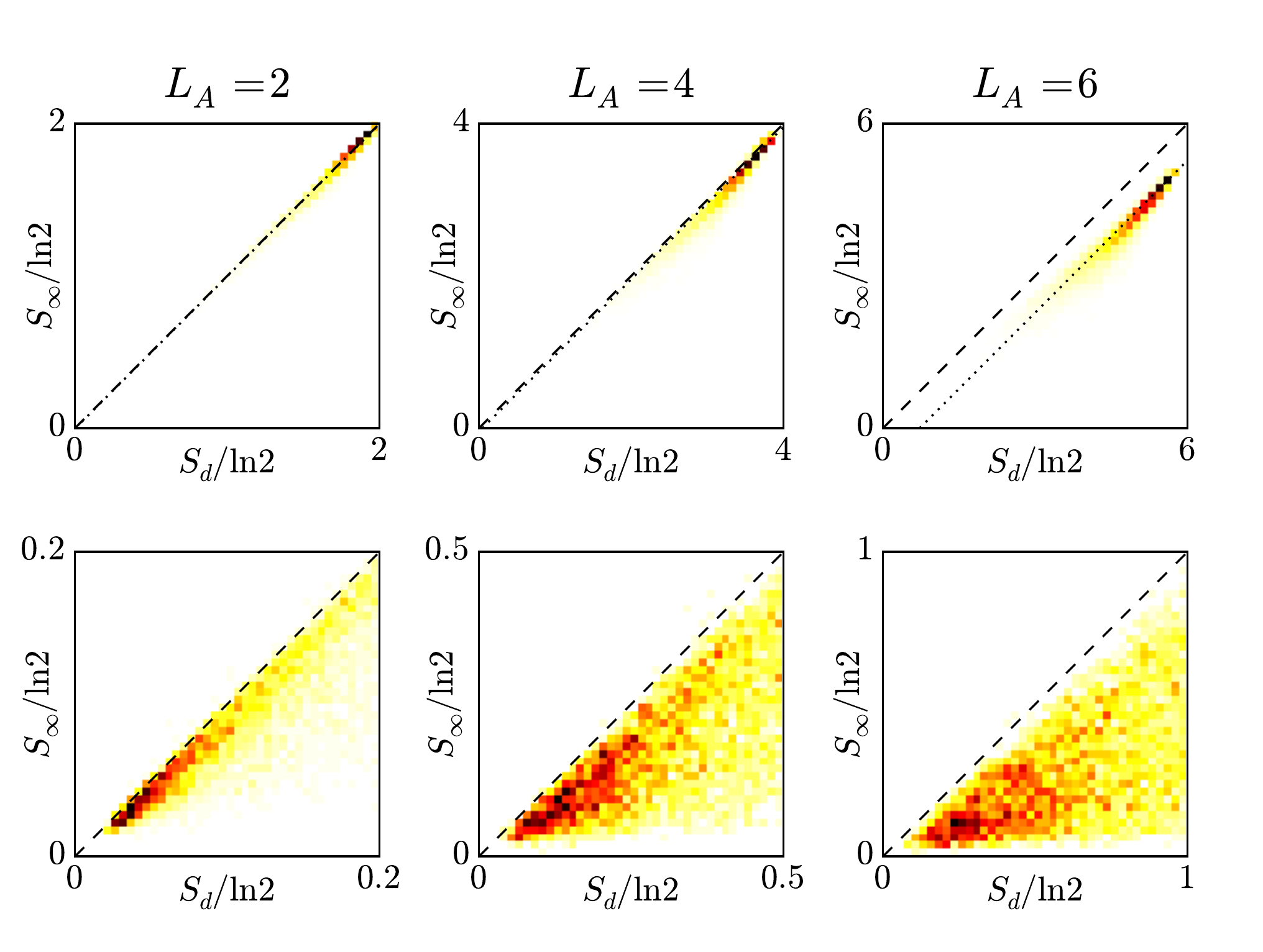}
\end{center}
\caption{(color online).
Correlation between entanglement at long times after the quench with diagonal ensemble for weak ($\eta=1$ in upper panel) and strong ($\eta=10$ in lower panel) disorder for $L=12$.
The entanglement is better predicted by the diagonal ensemble if the subsystem is much smaller than the total system.
The dotted lines in the upper panel include Page's correction $S_c$ and is given by $S_{\infty} = S_d - S_c$.
}
\label{SubsystemSize}
\end{figure}

MBL systems exhibit an emergent integrability in terms of local integrals of motion.~\cite{Huse2013a,Serbyn2013a,Imbrie2014}
As a result the thermalization picture breaks down, as one needs to take into account many conserved quantities, not just the energy.
The diagonal ensemble is suitable to handle such situations.~\cite{Rigol2008,Mondaini2015}
For a given initial state $|\psi_0 \rangle$ the diagonal ensemble density matrix is defined as
$$\rho_{_d} = \sum_i |\langle \psi_0|E_i\rangle|^2 |E_i \rangle \langle E_i|, $$
where $|E_i \rangle$'s are the energy eigenstates of the system.
Whereas for a given Hamiltonian the thermal state depends only on the energy of the initial state, the diagonal ensemble has more information.
By definition the diagonal ensemble is obtained by averaging the density matrix at all times, hence it can trivially estimate a very long time average of any physical observable.
However here we want to check whether it can predict the properties at long times.
There is a deficit in entropy when it is measured for an equal partition if the full system is in a pure state.\cite{Page1993,Garrison2015}
To avoid large deficit we use an unequal partition ($L_A=4, L_B=8$) while comparing asymptotic and diagonal ensemble properties in Fig.~\ref{DiagonalPrediction}.
We find that the diagonal ensemble makes an almost perfect prediction for bipartite fluctuations both in the weak and strong disorder case, while the prediction for entanglement is good only for weak disorder.
We present the effect of subsystem size in Fig.~\ref{SubsystemSize}, which shows that if the subsystem is much smaller than the total system the diagonal ensemble makes better prediction for entanglement.
For weak disorder the prediction is consistent when Page's correction $S_c = d_A / (d_A+d_B)$ is taken into account, $d_A$ and $d_B$ being the Hilbert space dimension of subsystems $A$ and $B$ respectively.~\cite{Page1993}

\subsection{Anderson localization} \label{AndersonSection}
\begin{figure}
\begin{center}
\includegraphics[width=0.99\linewidth]{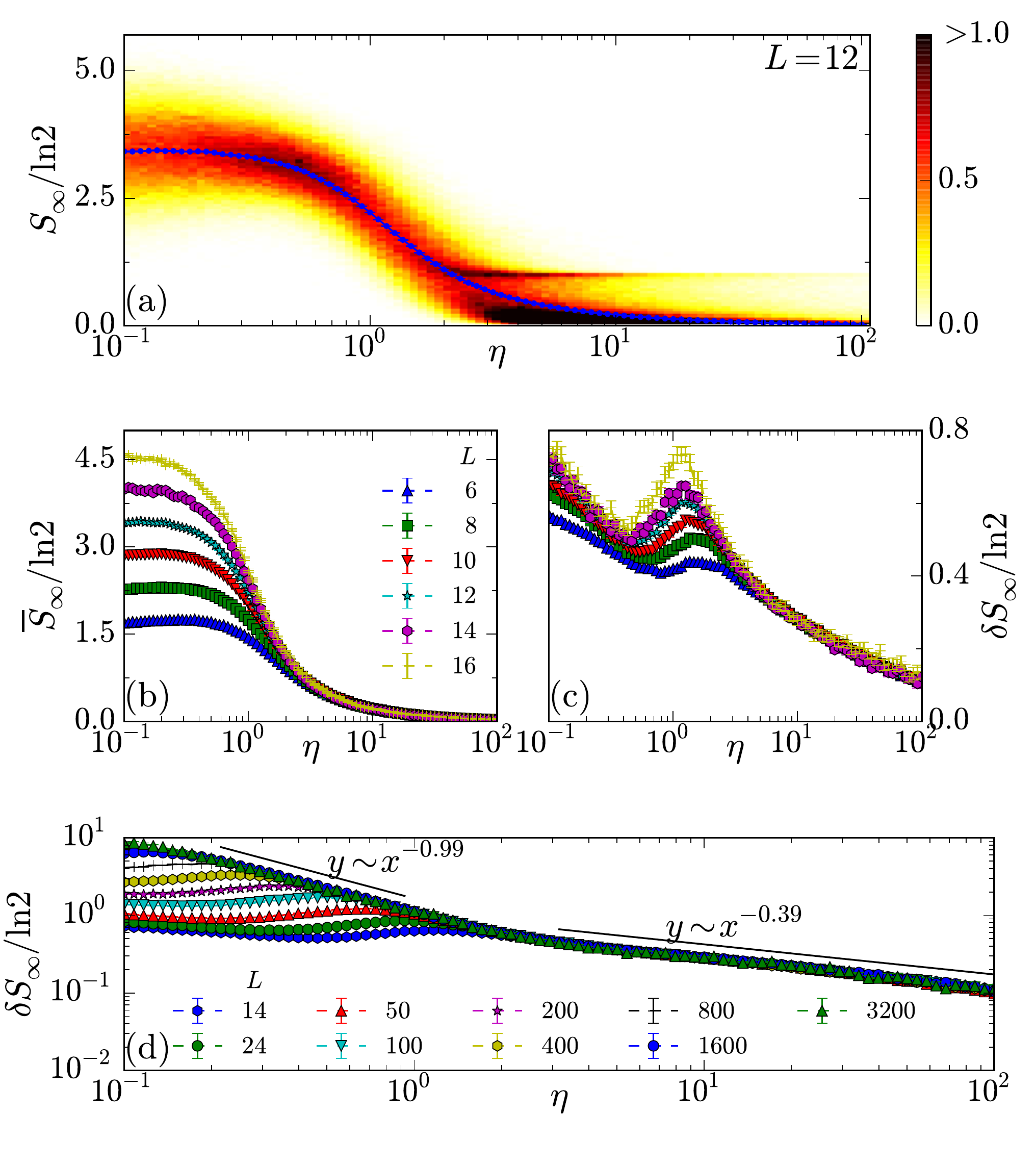}
\end{center}
\caption{(color online). Entanglement in the non-interacting system.
(a) Change in the distribution of asymptotic entanglement $S_{\infty}$ as a function of disorder ($\eta$).
Mean (b) and standard deviation (c) of $S_{\infty}$ versus $\eta$.
(d) Standard deviation of $S_{\infty}$ vs $\eta$ for large system size using free fermion simulation shows different scaling for weak and strong disorder.
}
\label{ent_sat_anderson}
\end{figure}

So far we have considered the quench properties of a disordered system in presence of interactions and the effects of localization.
However a non-interacting system ($\Delta=0$) localizes in the presence of an arbitrary weak uncorrelated disorder in 1D (Anderson localization).~\cite{Evers2008}
The localization length will be large at weak disorder and as a result there will be crossover-like behavior as the strength of disorder is increased for a finite system.
We perform global quench simulations of the non-interacting system and compare its asymptotic properties to that of the interacting system.
As noted in earlier studies~\cite{Bardarson2012,Serbyn2013} entanglement does not grow logarithmically for any disorder in this case but rather saturates after the initial rapid increase.
The asymptotic distribution shows a bimodal distribution beyond some disorder strength, see Fig.~\ref{ent_sat_anderson}~(a).
For small system sizes, the non-interacting system shows a crossover-like behavior and for very weak disorder it will effectively be in an extended phase as the localization length would be much larger than the system size.
However even in such situation, there is a qualitative difference between the interacting and non-interacting cases---the fluctuations in saturation value of entanglement caused by disorder are very strong for the non-interacting case (Fig.~\ref{ent_sat_anderson}) as compared to the interacting system~(Fig.~\ref{ent_sat}).
This is also manifested in the standard deviation of entanglement $\delta S_{\infty}$ which decreases as disorder goes to zero for the interacting system while the opposite behavior is observed for small systems in the non-interacting case, compare Figs.~\ref{ent_sat}~(c) and \ref{ent_sat_anderson}~(c).
This difference is seen even for the bipartite fluctuations and we expect a similar behavior for all observables.
Another important difference between the two cases is that the entanglement is independent of system size for strong disorder in the non-interacting system which is a consequence of the absence of logarithmic growth.
We confirm that the crossover-like behavior is indeed a finite size effect by simulating the quench dynamics in much larger systems~\cite{Peschel2003} and observing that the location of the local maxima in $\delta S_{\infty}$ shifts to lower values of $\eta$ as we increase the system size, see Fig.~\ref{ent_sat_anderson}~(d).
For a larger system size the single particle localization length becomes comparable to the system size at a weaker disorder.
We also note the appearance of different scaling for weak and strong disorder from the large system simulation of the non-interacting problem.
This different scaling appears to be the limiting behavior with increasing system size, i.e., it is independent of system size for large enough systems.
We speculate that this change in scaling is related to the localization length becoming comparable to lattice spacing.

\section{Summary and Conclusions} \label{Conclusions}
To summarize, we have studied the effect of disorder on global quench dynamics in a 1D spin chain and found that the asymptotic behavior of different physical quantities show signatures of the many-body localization transition.
We first reproduced the result that though the mean entanglement after very long time shows a volume law for both weak and strong disorder, the value itself decreases rapidly as the disorder is increased beyond some critical value.
More importantly the standard deviation of entanglement at large times behaves very similar to thermodynamic fluctuations and shows a peak near the transition.
We find similar behavior for bipartite fluctuations with one important difference, the bipartite fluctuations show an area law in the localized phase and its asymptotic values are independent of system size for strong disorder.
Near the transition the standard deviation of bipartite fluctuations also behave like thermodynamic fluctuations, a signature that can potentially be measured in cold atoms experiments using single atom microscopy.
We then compared the asymptotic properties following the quench with canonical ensemble at suitable temperature and explicitly showed the breakdown of thermalization in the localized phase.
This breakdown is a result of emergence of quasi-local integrals of motion in the system and we make very good predictions of the asymptotic bipartite fluctuations once these integrals are taken into account by using the diagonal ensemble.
We finally highlighted the effects of interactions by performing similar quench study of the non-interacting system.
We hope that this work would lead to more experimental efforts to study the MBL phenomena using particle number fluctuations.

\section*{Acknowledgements}
We thank M. Rigol, A. Polkovnikov, F. Essler and T. Grover for helpful discussions.
\bibliography{bibliography}
\end{document}